\documentclass[10pt]{article}
\newcommand{\dd}{\mbox{\rm d}}
\newcommand{\gam}{\gamma}

\newcommand{\oo}{\over}
\newcommand{\p}{\partial}

\newcommand{\be}{\begin{equation}}
\newcommand{\bear}{\begin{eqnarray}}
\newcommand{\ear}{\end{eqnarray}}
\newcommand{\ee}{\end{equation}}
\newcommand{\lbl}{\label}
\newcommand{\bi}{\bibitem}
\newcommand{\ci}{\cite}

\newcommand{\vs}{\vspace}

\begin{document}

\ 

\rightline{Report IJS/TP-22/04}

\vs{20mm}

\begin{center}

{\bf CLIFFORD SPACE AS A GENERALIZATION OF SPACETIME:}

\vs{2mm}

{\bf PROSPECTS FOR UNIFICATION IN PHYSICS}\footnote{Talk presented
at {\it 4th Vigier Symposium: The Search For Unity in Physics},
September 15th--19th, 2003, Paris, France.}

\vs{3mm}

Matej Pav\v si\v c

Jo\v zef Stefan Institute, Jamova 39,
1000 Ljubljana, Slovenia

e-mail: matej.pavsic@ijs.si

\end{center}

\begin{abstract}

The geometric calculus based on Clifford algebra is a very useful
tool for geometry and physics. It describes a geometric structure
which is much richer than the ordinary geometry of spacetime.
A Clifford manifold ($C$-space) consists not only of points, but also
of 1-loops, 2-loops, etc.. They are associated with multivectors which
are the wedge product of the basis vectors, the generators of Clifford
algebra. We assume that $C$-space is the true space in which physics
takes place and that physical quantities are Clifford algebra valued
objects, namely, superpositions of multivectors, called  Clifford
aggregates or polyvectors. We explore some very promising features
of physics in Clifford space, in particular those related to a consistent
construction of string theory and quantum field theory.

\end{abstract}

\vs{5mm}

{\bf 1. Introduction}

\vs{2mm}

In recent years there has been the interest in considering a theory
in which spacetime is replaced by Clifford space 
\ci{Pezzaglia}--\ci{CliffConform}. The latter
space describes not only   points, but also areas, volumes, etc., altogether
on the same footing. In refs. \ci{PavsicArena} it was proposed that
the coordinates of Clifford space can be
interpreted as generalizing the concept of center of mass coordinates
of extended objects. In other words, the extended objects can be
modeled by coordinates of Clifford space.

Instead of the usual theory of relativity in spacetime $M_n$
we have the relativity in $C$-space. The latter space
has dimension $2^n$ and signature $(+++ ... --- ...)$, where the number of
plus and minus signs is the same, namely $2^n/2$. This has
consequences for string theory which can be formulated without
central terms in Virasoro algebra even when the dimension of the
underlying spacetime is four, provided that the Jackiw definition \ci{Jackiw}
of vacuum is employed \ci{PseudoHarm}. We do not
need a higher dimensional target spacetime for a consistent formulation
of (quantized) string theory. Instead of a higher dimensional space
we have Clifford space which also provides a natural framework 
\ci{Pavsic,PavsicBook} for
description of superstrings and supersymmetry, since spinors are just 
the elements of left or right minimal ideals of Clifford algebra
\ci{Teitler}--\ci{Mankoc}.

When considering field theory and assuming the Jackiw definition
of vacuum state, the concept of $C$-space enables a formulation in which
zero point energies belonging to positive and negative signature
degrees of freedom cancel out \ci{PseudoHarm}, while preserving
the Casimir effect.
This provides a resolution of the cosmological constant problem.

Instead of flat $C$-space we may consider a curved $C$-space. As the
passage from flat Minkowski spacetime to curved spacetime had
provided us with a tremendous insight into the nature of one
of the fundamental interactions, namely gravity, so the introduction
of a curved $C$-space will presumably even further increase our
understanding of the other fundamental interactions and their unification
with gravity.

\vs{5mm}

{\bf 2. On the Relativity in Clifford space}

\vs{2mm}

Let us start with $n$-dimensional flat spacetime $M_n$ whose signature
is $(+ - - - - ...)$. A point $P$ in $M_n$ can be associated with a {\it
vector} $x$ joining the coordinate origin $O$ and $P$. Vectors can
be represented by Clifford numbers according to
\be
    x = x^\mu \gamma_\mu
\lbl{2.1}
\ee
where $\gamma_\mu$ satisfy the Clifford algebra relation
\be
     \gamma_\mu \cdot \gamma_\nu \equiv {1\oo 2} (\gamma_\mu
    \gamma_\nu + \gamma_\nu \gamma_\mu) = g_{\mu \nu}
\lbl{2.2}
\ee
and $g_{\mu \nu}$ is the metric of $M_n$.

The latter relation is the {\it inner product} between two basis vectors
$\gamma_\mu$ and $\gamma_\nu$. It is just the {\it symmetric part}
of the Clifford product $\gamma_\mu \gamma_\nu$.

Besides vectors we can consider {\it bivectors} defined by the wedge product
$\gam_\mu \wedge \gam_\nu$ which is the {\it antisymmetric part}
of $\gam_\mu \wedge \gam_\nu$:
\be
    \gam_\mu \wedge \gam_\nu = {1\oo 2} (\gam_\mu \gam_\nu - \gam_\nu \gam_\mu)
\equiv {1\oo 2} [\gam_\mu, \gam_\nu]
\lbl{2.3}
\ee
and in general multivectors or $r$-vectors
\be
     \gam_{\mu_1}\wedge \gam_{\mu_2} \wedge ... \wedge \gam_{\mu_r}
 = {1\oo r!} [\gam_{\mu_1},\gam_{\mu_2},...,\gam_{\mu_r}]
\lbl{2.4}
\ee
Bivectors, trivectors, etc., represent oriented areas, volumes, etc., respectively.
The highest grade $r$ in $M_n$ is $r=n$. The $r$-vectors of $r>n$ are
identically zero.

A generic {\it Clifford number} (called also {\it polyvector} of {\it
Clifford aggregate}) is
\be
    X = \sigma {\underline 1} + x^\mu \gam_\mu + {1\oo 2} x^{\mu_1 \mu_2}
   \gam_{\mu_1 \mu_1} + ... + {1\oo n!} x^{\mu_1 ... \mu_n}
   \gam_{\mu_1 ... \mu_n} \equiv x^M \gam_M
\lbl{2.5}
\ee
where $\gam_{\mu_1 ... \mu_r} \equiv \gam_{\mu_1} \wedge \gam_{\mu_2}
\wedge ... \wedge \gam_{\mu_r}$ and 
\bear
    &&x^M = (\sigma, x^{\mu_1}, x^{\mu_1 \mu_2},..., 
  x^{\mu_1 ... \mu_r} )\nonumber \\
   &&\gam_M = ({\underline 1}, \gam_\mu , \gam_{\mu_1 \mu_2},...,
\gam_{\mu_1 ... \mu_r}) \; , \quad \mu_1 < \mu_2 < ... < \mu_r 
\lbl{2.6}
\ear
are respectively coordinates and basis elements of Clifford algebra.

The coordinates $x^{\mu_1 ... \mu_r}$ determine an oriented
$r$-area. They say nothing about the precise form of the $(r-1)$-loop
enclosing the $r$-area. The coordinates 
$\sigma, ~x^{\mu_1},~ x^{\mu_1 \mu_2},...$ provide a
means for a description of extended objects. If an objects is extended,
then not only its center of mass coordinates $x^\mu$, but also the
higher grade coordinates $ x^{\mu_1 \mu_2},~ x^{\mu_1 \mu_2, \mu_3}$,...,
associated with the object extension, are different from zero, in general.
Those higher grade coordinates model the extended object.
A detailed description is provided in ref. \ci{PavsicArena}.

Since $x^M$ assume any real value, the set of all possible
$X$ forms a $2^n$-dimensional manifold, called {\it Clifford space},
or shortly {\it $C$-space}.

Let us define the quadratic form by means of the scalar product
\be
    |\dd X|^2 \equiv \dd X^{\ddagger} * \dd X = \dd x^M \dd x^N
   G_{MN} \equiv \dd x^M \dd x_M
\lbl{2.7}
\ee
where the metric of $C$-space is given by
\be
      G_{MN} = \gam_M^{\ddagger} * \gam_N
\lbl{2.8}
\ee
The operation $\ddagger$ reverses the order of vectors:
\be
    (\gam_{\mu_1} \gam_{\mu_2}...\gam_{\mu_r})^{\ddagger}
  = \gam_{\mu_r} ... \gam_{\mu_2} \gam_{\mu_1}
\lbl{2.8a}
\ee
Indices are lowered and raised by $G_{MN}$ and its inverse
$G^{MN}$, respectively. The following relation is satisfied:
 \be
     G^{MJ} G_{JN} = {\delta^M}_N
\lbl{2.9}
\ee

Eq.(\ref{2.7}) is the expression for the {\it line element} in
$C$-space. If $C$-space is generated from the basis vectors
$\gam_\mu$ of spacetime $M_n$ with signature
$(+ - - - - - ...)$, then the signature of $C$-space is
$(+ + + ... - - -...)$, where the nuber of plus and minus signs is
the same, namely, $2^n/2$. This has some important
consequences that we are going to investigate.

We assume that $2^n$-dimensional Clifford space is the
arena in which physics takes place. We can take $n=4$,
so that the spacetime from which we start is just the
4-dimensional Minkowski space $M_4$. The corresponding
Clifford space has then 16 dimensions. In $C$-space the
usual points, lines, surfaces, volumes and 4-volumes are
all described on the same footing and can be transformed
into each other by rotations in $C$-space (called polydimensional
rotations):
\be
    x'^M = {L^M}_N x^N
\lbl{2.10}
\ee
subjected to the condition $|\dd X'|^2 = |\dd X|^2$.

We can now envisage that physical objects are described
in terms of $x^M = (\sigma, x^\mu, x^{\mu \nu},...)$. The
first straightforward possibility is to introduce a single
parameter $\tau$ and consider a mapping
\be
    \tau \rightarrow x^M = X^M (\tau)
\lbl{2.10a}
\ee
where $X^M (\tau)$ are 16 embedding functions that
describe a worldline in $C$-space. From the point of view of
$C$-space, $X^M (\tau)$ describe a wordlline of a ``point
particle": At every value of $\tau$ we have a {\it point} in
$C$-space. But from the perspective of the underlying
4-dimensional spacetime, $X^M (\tau)$ describe an extended
object, sampled by the center of mass coordinates $X^\mu (\tau)$
and the coordinates
$X^{\mu_1 \mu_2}(\tau),..., X^{\mu_1 \mu_2 \mu_3 \mu_4} (\tau)$.
They are a generalization of the center of mass coordinates in the sense
that they provide information about the object 2-vector, 3-vector, and
4-vector extension and orientation\footnote{A systematic and detailed
treatment is in ref. \ci{PavsicArena}.}.

The dynamics of such an object is determined by the action
\be
    I[X] = M \int \dd \tau \, ({\dot X}^{\ddagger} *{\dot X})^{1\oo 2} =
   M \int \dd \tau ({\dot X}^M {\dot X}_M)^{1\oo 2}
\lbl{2.11}
\ee
The dynamical variables are given by the polyvector
\be
   X = X^M \gam_M = \sigma {\underline 1} + X^\mu \gam_\mu +
  X^{\mu_1 \mu_2} \gam_{\mu_1 \mu_2} + ... X^{\mu_1 ...\mu_n} \gam_{\mu_1 ... \mu_n}
\lbl{2.11a}
\ee
whilst
\be 
{\dot X} = {\dot X}^M \gam_M = {\dot \sigma} {\underline 1} + {\dot X}^\mu \gam_\mu +
  {\dot X}^{\mu_1 \mu_2} \gam_{\mu_1 \mu_2} + ... 
 {\dot X}^{\mu_1 ...\mu_n} \gam_{\mu_1 ... \mu_n}
\lbl{2.11b}
\ee
is the velocity polyvector, where ${\dot X}^M \equiv \dd X^M/\dd \tau$.

In the action (\ref{2.11}) we have a straightforward generalization of the
relativistic point particle in $M_4$:
\be
   I[X^\mu] = m \int \dd \tau ({\dot X}^\mu {\dot X}_\mu)^{1\oo 2}
  \; , \quad \mu = 0,1,2,3
\lbl{2.12}
\ee
 If a particle is extended, then (\ref{2.12}) provides only a very incomplete
description. A more complete description is given by the action
(\ref{2.11}), in which the $C$-space embedding functions $X^M (\tau)$
sample the objects extension.

\vs{5mm}

{\bf 3. Strings and Clifford space}

\vs{2mm}

Usual strings are described by the mapping $(\tau,\sigma) \rightarrow
x^\mu = X^\mu (\tau, \sigma)$, where the embedding functions
$X^\mu (\tau,\sigma)$ describe a 2-dimensional worldsheet swept
by a string. The action is given by the requirement that the area of the
worldsheet be ``minimal" (extremal). Such action is invariant under
reparametrizations of $(\tau,\sigma)$. There are several equivalent
forms of the action including the ``$\sigma$-model action" which, in
the conformal gauge, can be written as
\be
   I[X^\mu] = {\kappa\oo 2} \int \dd \tau \, \dd \sigma \, ({\dot X}^\mu
   {\dot X}_\mu - X'^\mu X'_\mu)
\lbl{3.1}
\ee
where ${\dot X}^\mu \equiv \dd X^\mu/\dd \tau$ and $X'^\mu \equiv
\dd X^\mu/\dd \sigma$. Here $\kappa$ is the string tension, usually
written as $\kappa = 1/(2 \pi \alpha')$.

String coordinates $X^\mu$ and momenta $P_\mu = \p L/\p {\dot X}^\mu
=\kappa {\dot X}_\mu$
satisfy the following constraints $(\sigma \in [0,\pi])$:
\be
      \varphi_1 (\sigma) = P^\mu P_\mu + {{X'^\mu X'_\mu}\oo {(2 \pi 
  \alpha')^2}} \approx 0 \; \qquad \varphi_2 (\sigma) =
   {{P^\mu X'_\mu}\oo {\pi \alpha'}} \approx 0
\lbl{3.1a}
\ee
which can be written as a single constraint on the interval $\sigma \in
[- \pi, \pi]$
\be
     \Pi^\mu \Pi_\mu (\sigma) \approx 0 \; \qquad 
     \Pi^\mu = P^\mu + {X'^\mu\oo 2 \pi \alpha'}
\lbl{3.1b}
\ee     
to which the open string is symmetrically extended.
(For more details see the literature on strings, e.g., \ci{Strings}.)

If we generalize the action (\ref{3.1}) to $C$-space, we obtain
\be
     I[X] = {\kappa\oo 2} \int \dd \tau \, \dd \sigma \, ({\dot X}^M {\dot X}^N-
  X'^M X'_N)G_{MN}
\lbl{3.2}
\ee
where $\kappa$ is the generalized string tension.
Taking 4-dimensional spacetime, there are $D=2^4 = 16$ dimensions of $C$-space.
Its signature $(+++...- - - ...)$ has 8 plus and 8 minus signs.

Let  us consider the case of open string satisfying the boundary condition
$X'^M = 0$ at $\sigma = 0$ and $\sigma = \pi$. Then we can make
the expansion
\be
      X^M (\tau,\sigma) = \sum_{n=-\infty}^\infty X_n^M (\tau)
      \, {\rm e}^{i n \sigma}
\lbl{3.3}
\ee
where from the reality condition $(X^M)^* = X^M$ it follows
\be
     X_n^M = X_{-n}^M
\lbl{3.4}
\ee
Inserting (\ref{3.3}) into (\ref{3.2}), integrating over $\sigma$ and taking
into account (\ref{3.4}) we obtain the action expressed in terms of
$X_n^M (\tau)$:
\be
   I[X_n^M] = {\kappa'\oo 2} \int \dd \tau \, \sum_{n=-\infty}^\infty
     ({\dot X}_n^M {\dot X}_n^N - n^2 X_n^M X_n^N)G_{MN}
\lbl{3.5}     
\ee
where $\kappa' = 2 \pi \kappa = 1/\alpha'$. This is just the action of infinite number of
harmonic oscillators.

The Hamiltonian corresponding to the action (\ref{3.5}) is
\be
    H = {1\oo 2} \sum_{n=-\infty}^{\infty}  \left ( {1\oo \kappa'}
    P_n^M P_{nM} + \kappa' \, n^2 X_n^M X_{nM} \right )
\lbl{3.6}
\ee
Let us introduce
\bear
     &&a_n^M = {1\oo \sqrt{2}} \left ( {1\oo \sqrt{\kappa'}} P_n^M
     - i n \sqrt{\kappa'} \, X_n \right ) \nonumber \\
     &&{a_n^M}^\dagger = {1\oo \sqrt{2}} \left ( {1\oo \sqrt{\kappa'}} P_n^M
     + i n \sqrt{\kappa'} \, X_n \right )
\lbl{3.7}
\ear
We see that $a_{-n}^M = {a_N^M}^\dagger$.
Rewriting $H$ in terms of $a_n^M$, ${a_n^M}^\dagger$ we obtain
\be
    H = {1\oo 2} \sum_{n=-\infty}^\infty ({a_n^M}^\dagger a_{nM} +
    a_{nM} {a_n^M}^\dagger ) = \sum_{n=1}^{\infty}
    ({a_n^M}^\dagger a_{nM} + a_{nM} {a_n^M}^\dagger )
    + {1\oo {2 \kappa'}} P_0^M P_{0 M}
\lbl{3.8}
\ee

Upon quantization we have
\be
     [X_n^M,P_{nN}] = i {\delta^M}_N \quad {\rm or} \quad 
     [X_n^M,P_n^N] = i G^{MN}
\lbl{3.9}
\ee
and
\be
     [a_n^M,a_{nN}^\dagger] = i {\delta^M}_N \quad {\rm or} \quad 
     [a_n^M,{a_n^N}^\dagger] = G^{MN}
\lbl{3.10}
\ee

In order to construct the Fock space of excited states, one has first
to define a {\it vacuum state}. There are two possible choices
\ci{PseudoHarm}.

{\it Possibility I}. Conventionally, vacuum state is defined
according to
\be
      a_n^M|0 \rangle = 0 \; , \qquad ~~ n \ge 1
\lbl{3.11}
\ee
and the excited part of the Hamiltonian
$H_{\rm exc} = H-(1/\kappa') P_0^M P_{0M}$, after using (\ref{3.10}) and 
(\ref{3.11}) is
\be
    H_{\rm exc} = \sum_{n=-\infty}^\infty ({a_n^M}^\dagger a_{nM} +{D\oo 2})
      = 2  \sum_{n=1}^\infty ({a_n^M}^\dagger a_{nM} +{D\oo 2}) 
\lbl{3. 12}
\ee
$$D= {\delta^M}_M = G^{MN} G_{MN}$$
Its eigenvalues are all positive\footnote{This is so even for those components
$a_n^M$ that belong to negative signature: negative sign of a term in
${a_n^M}^\dagger a_{nM}$ is compensated by negative sign in the commutation
relation (\ref{3.10}).} and there is the non vanishing zero point
energy. But there exist negative norm states.

{\it Possibility II}.  Let us split $a_n^M = (a_n^A, a_n^{\bar A})$ where
the indices $A,~{\bar A}$ refer to the components with positive and negative
signature, respectively, and let us define vacuum according to
\be
    a_n^A |0 \rangle = 0 \; , \qquad {(a_n^{\bar A})}^\dagger |0 \rangle = 0
    \; , \quad n \ge 1
\lbl{3.13}
\ee
Using (\ref{3.10}) we obtain the following Hamlitonian, in which
annihilation operators, defined in eq.\,(\ref{3.13}), are on the right:
\be
    H_{\rm exc} = 2 \sum_{n=1}^\infty ({a_n^A}^\dagger a_{nA} + {R\oo 2}
       + a_{n{\bar A}} {a_n^{\bar A}}^\dagger - {S\oo 2})
\lbl{3.14}
\ee
where $R = {\delta_A}^A$ and $S= {\delta _{\bar A}}^{\bar A}$.
{\it There are no negative norm states.}

If the number of positive and negative signature components is the
same, i.e., $R=S$, then the above Hamiltonian has vanishing zero point
energy:
\be H_{\rm exc} = 2 \sum_{n=1}^\infty ({a_n^A}^\dagger a_{nA} 
       + a_{n{\bar A}} {a_n^{\bar A}}^\dagger)
\lbl{3.15}
\ee
Its eigenvalues can be positive or negative, depending on which components
(positive or negative signature) are excited.

An immediate objection could arise at this point, namely, that since
the spectrum of the Hamiltonian is not bounded from below, the system
described by $H$ of eq.(\ref{3.14}) or (\ref{3.15}) is unstable. This
objection would only hold if the kinetic terms ${\dot X}_n^M {\dot X}_{nM}$
in the action (\ref{3.5}) (or the terms $P_n^M p_{nM}$ in the Hamiltonian
(\ref{3.6})) were all positive, so that negative eigenvalues of $H$ would come
from the negative potential terms in $n^2 X_n^M X_{nM}$. But since our
metric is pseudo-Euclidean, whenever a term in the potential is negative,
also the corresponding kinetic term is negative. Therefore, the
acceleration corresponding to negative signature term is proportional to
the {\it plus} gradient of potential (and not to the minus gradient of
potential as it is the case for positive signature term); such system is
{\it stable} if potential has {\it maximum}, i.e., if it has an upper bound
(and not a lower bound). The overall change
of sign of the action (Lagrangian) has no influence on the equations of
motion (and thus on stability).

In the bosonic string theory based on the ordinary definition of
vacuum (Possibility I) and formulated in $D$-dimensional spacetime
with signature $(+ - - - ... - - -)$ there are negative norm states,
unless $D=26$. Consistency of the string theory requires extra dimensions,
besides the usual four dimensions of spacetime.

My proposal is that, instead of adding extra dimensions to spacetime,
we can start from 4-dimensional spacetime $M_4$ with signature
$(+ - - -)$ and consider the Clifford space ${\cal C}_{M_4}$ ($C$-space)
whose dimension is 16 and signature $(8+,8-)$. {\it The necessary extra
dimensions for consistency of string theory are in $C$-space.} This
also automatically brings {\it spinors} into the game. It is an old
observation that spinors are the elements of left or right ideals of
Clifford algebras \ci{Teitler}--\ci{Lounesto} (see also a very lucid and
systematic recent exposition in refs.\,\ci{Mankoc}). In other words, spinors
are particular sort of
polyvectors \ci{PavsicBook}. Therefore, the string coordinate
polyvectors contain spinors.
This is an alternative way of introducing spinors into the string theory
\ci{PavsicBook,Castro-Pavsic}.
           
Let the constraints (\ref{3.1a}),(\ref{3.1b}) be generalized to $C$-space.
So we obtain
\be
    \Pi^M \Pi_M \approx 0 \; , \qquad \Pi^M = P^M + {X'^M\oo 2 \pi \alpha'}
\lbl{3.16}
\ee
Using (\ref{3.3}) and expanding the momentum $P^M (\sigma)$ according to
\be
     P^M = \sum_{n=-\infty}^\infty P_n^M {\rm e}^{i n \sigma}
\lbl{3.17}
\ee
we can calculate the Fourier coefficients of the constraint (called Virasoro
generators):
\be
    L_n = {\pi \alpha'\oo 2} \int_{-\pi}^\pi \dd \sigma \, {\rm e}^{-in \sigma}
    \, \Pi^M \Pi_M = {1\oo 2} \sum_{r=-\infty}^\infty a_n^M a_{r-n}^N G_{MN}
\lbl{3.18}
\ee

If we calculate the commutators  of Virasoro generators, then, after
putting to the left those operators which according to the
vacuum definition (\ref{3.13}) act as creation operators, 
we obtain the following relation
\be
    [L_m, L_n] = (m-n) L_{n+m}
\lbl{3.19}
\ee
in which there are no central terms. The terms which arise after
reordering of operators have opposite signs for positive and negative
signature components,
and thus cancel out. The algebra of Virasoro generators is thus closed,
which automatically assures consistency of quantum string theory.

\vs{5mm}

{\bf 4. Quantum Field Theory and $C$-space}

\vs{2mm}

In previous section we considered string theory which can be considered
as a theory of $D$ fields $X^M (\tau, \sigma)$ over a 2-dimensional
space. Let us now turn to the theory of $n$ scalar fields $\phi^a (x^\mu)$,
$a=0,1,2,...,n-1$, over the 4-dimensional spacetime parametrized by
coordinates $x^\mu$, $\mu = 0,1,2,3$. The action for such system is
\be
    I[\phi^a] = {1\oo 2} \int \dd^4 x \, \sqrt{-g} (g^{\mu \nu}
    \p_\mu \phi^a \, \p_\nu \phi^b - m^2 \phi^a \phi^b ) \gamma_{ab}
\lbl{4.1}
\ee
where $\gam_{ab}$ is the metric in the space of fields $\phi^a$,
and $g_{\mu \nu}$ the metric of spacetime. Let us assume that $g_{\mu \nu}
=\eta_{\mu \nu}$ is the metric of flat spacetime

The canonical momenta are
\be
     \pi_a = {\p {\cal L}\oo \p \p_0 \phi^a} = \p^0 \phi_a = \p_0 \phi_a \equiv
     {\dot \phi}_a
\lbl{4.2}
\ee

Upon quantization the following equal time commutation relations are
satisfied:
\be
    [\phi^a ({\bf x}),\pi_b ({\bf x}')] = i \delta^3 ({\bf x} - {\bf x}')
    {\delta^a}_b
\lbl{4.3}
\ee
The Hamiltonian is given by
\be
    H= {1\oo 2} \int \dd^3 x \, ( {\dot \phi}^a {\dot \phi}^b -
    \p_i \phi^a \p^i \phi^b + m^2 \phi^a \phi^b) \gam_{ab}
\lbl{4.4}
\ee
where $i= 1,2,...,n-1$. We shall assume that $\gam_{ab}$ is diagonal.
A general solution to the equations of motion derived from the action
(\ref{4.1}) can be written in the form
\be
    \phi^a = \int {{\dd^3 {\bf k}}\oo {(2 \pi)^3}} \, 
    {1\oo {2 \omega_{\bf k}}} (a^a ({\bf k}) {\rm e}^{-ikx} + {a^a}^\dagger
    ({\bf k}) {\rm e}^{ikx} )
\lbl{4.5}
\ee
Here\footnote{We use units in which $\hbar = c = 1$.} $\omega_{\bf k} 
\equiv |\sqrt{m^2 + {\bf k}^2}|$. The creation and annihilation operators
satisfy the commutation relations
\be
    [a^a ({\bf k}),{a_b}^\dagger ({\bf k})] = (2 \pi)^3 \, 2 \omega_{\bf k}
    \, \delta^3 ({\bf k} - {\bf k}') {\delta^a}_b
\lbl{4.5a}
\ee
or
\be
    [a^a ({\bf k}),{a^b}^\dagger ({\bf k})] = (2 \pi)^3 \, 2 \omega_{\bf k}
    \, \delta^3 ({\bf k} - {\bf k}') \gam^{ab}
\lbl{4.5b}
\ee   
Inserting the expansion (\ref{4.5}) of fields $\phi^a$ into the Hamiltonian
(\ref{4.4}) we obtain
\be
    H = {1\oo 2} \int {{\dd^3 {\bf k}}\oo {(2 \pi)^3}} \, 
    {\omega_{\bf k}\oo {2 \omega_{\bf k}}} ({a^a}^\dagger ({\bf k})
    a^b ({\bf k}) + a^a ({\bf k}) {a^b}^\dagger ({\bf k}) ) \gam_{ab}
\lbl{4.6}
\ee

Let us assume that the signature of the metric $\gam_{ab}$ is pseudo-Euclidean,
and let us write
\be
    a^a ({\bf k}) = (a^\alpha, a^{\bar \alpha})
\lbl{4.7}
\ee
where $\alpha$ denotes positive and ${\bar \alpha}$ negative signature
components.

We will define vacuum according to
\be
    a^\alpha ({\bf k}) |0 \rangle 
     = 0 \; , \quad {a^{\bar \alpha}}^\dagger ({\bf k})
    |0 \rangle = 0
\lbl{4.8}
\ee
If we reorder the operators in the Hamiltonian (\ref{4.6}) so that the
annihilation operators (with respect to the vacuum definition
(\ref{4.8})) are on the right, we find
\be
    H = \int {{\dd^3 {\bf k}}\oo {(2 \pi)^3}} \, 
    {\omega_{\bf k}\oo {2 \omega_{\bf k}}} ({a^\alpha}^\dagger ({\bf k})
    a_\alpha ({\bf k}) + a^{\bar \alpha} ({\bf k}) 
    {a_{\bar \alpha}}^\dagger ({\bf k}) + {r\oo 2}  - {s\oo 2} ) 
\lbl{4.9}
\ee
where $r={\delta^\alpha}_\alpha$ and $s= {\delta^{\bar \alpha}}_{\bar \alpha}$.
In the case in which the signature has equal number of plus and minus signs,
i.e., when $r=s$, the zero point energies cancel out from the
Hamiltonian.

Now a question arises as to why should the space of fields have the metric
with $r=s$. Isn't it an ad hoc assumption?  As in the case of string
we can consider the space of fields $V_n$ just as a starting space,
with basis $e_a,~a=0,1,2,...,n-1$, from which we generate the
$2^n$-dimensional Clifford space ${\cal C}_{V_n}$ with basis
$e_A = ({\underline 1},e_a, e_{a_1 a_2},...,e_{a_1...a_n}),~a_1<a_2<...<a_n$.
If $V_n$ is a Euclidean space so that $e_a \cdot e_b = \delta_{ab}$ is
the Euclidean metric, then also the metric ${e_A}^\ddagger *e_B$ of
${\cal C}_{V_n}$ is Eucliddean. But, as it was pointed out in refs. 
\ci{Pavsic,PavsicBook},
instead of the basis $e_A$ we can take another basis, e.g.,
\be
    \gamma_A = ({\underline 1},\gam_a,\gam_{a_1a_2},...,\gam_{a1...a_n})
\lbl{4.10}
\ee
generated from the set of Clifford numbers $\gam_a = (e_0,\, e_i e_0),~
a=0,1,2,...,n-1;~i=1,2,...,n$ satisfying
\be
    \gam_a \cdot \gam_b \equiv {1\oo 2} (\gam_a \gam_b + \gam_b \gam_a) = 
    \eta_{ab}
\lbl{4.11}
\ee
where
\be
     \gam_{a_1 a_2 ... a_r} \equiv \gam_{a_1} \wedge \gam_{a2} \wedge ...
     \wedge \gam_{a_r} \equiv {1\oo 2} [\gam_{a1},\gam_{a2},..., \gam_{a_r}]
\lbl{4.12}
\ee
The metric
\be     {\gam_A}^\ddagger * \gam_B = G_{AB}
\lbl{4.13}
\ee
defined with respect to the new basis is pseudo-Euclidean, its signature
having $2^n/2$ plus and $2^n/2$ minus signs. For simplicity
we use in eq.(\ref{4.13}) the same symbol `$\ddagger$`, but now it
denotes reversion of new basis vectors: 
$(\gamma_{a_1}...\gam_{a_r})^{\ddagger} = \gam_{a_r} ...\gam_{a_1}$.

We assume that a field theory should be formulated in $C$-space. Instead
of the action (\ref{4.1}) we thus consider its generalization to $C$-space:
\be
     I = {1\oo 2} \int \dd^4 x \, \sqrt{-g} (g^{\mu \nu}
     \p_\mu \phi^A \p_\nu \phi^B - m^2 \phi^A \phi^B) G_{AB}
\lbl{4.14}
\ee
Here $\Phi=\phi^A \gam_A$ is a polyvector field and $\Phi^\ddagger*\Phi
= \phi^A \phi^B\, G_{AB}$. Since the metric $G_{AB}$ has
signature $(+++ ... - - - ...) = (R+, S-)$ with $R=S$, zero point energies
of a system based on the action (\ref{4.14}) cancel out: Vacuum energy
vanishes. Consequently, in such a theory there is no cosmological problem
\ci{PseudoHarm}. The small cosmological constant, as recently observed,
could be a residual effect of something else.

Cancellation of vacuum energies in the theory does not exclude 
\ci{PseudoHarm} the existence of well known effects, such as Casimir effect,
which are manifestation of vacuum energies.

\vs{5mm}

{\bf 5. Discussion and Conclusion}

\vs{2mm}

We started from 4-dimensionl spacetime $M_4$ and generalized it
to Clifford space ${\cal C}_{M_4}$. We considered a theory
of a 1-dimensional worldline (swept by  point particle) and of
a 2-dimensional worldsheet (swept by string) living in 16-dimensional
Clifford space. In this theory no extra dimensions of the
target spacetime are required. The necessary extra dimensions are in
Clifford space ($C$-space) generated by 4 independent basis vectors
$\gam_\mu$ of spacetime. So we obtain a framework in which fermions
(as elements of the minimal ideals of ${\cal C}_{M_4}$) also enter
the game. A next logical step is to generalize the $(p+1)$-dimensional
world manifold (including, for $p=1$, the string worldsheet $V_2$) to
to the corresponding Clifford space ${\cal C}_{V_{p+1}}$. Such theory
is discussed in ref. \ci{PavsicBook}.

We then consider the quantum theory of $n$ scalar fields and take the
space of fields as a starting space from which we construct the
corresponding Clifford space ${\cal C}_{V_n}$. Amongst available
$2^n$ basis elements of ${\cal C}_{V_n}$ we are free to choose $n$
elements, denoted $\gam_a,~a=0,1,2,...,n$, such that the inner products
defined according to  eq.(\ref{4.11}) form the Minkowski metric
with signature $(+ - - - ...)$. Taking $\gam_a$ as generators of Clifford
algebra and define the metric of $C$-space according to eq. (\ref{4.13}),
we obtain  that the signature of Clifford space is $(+++...- - - ...)$ with
$2^n$ plus and $2^n$ minus signs. We then find that the vacuum energy
belonging to the negative signature degrees of freedom cancel out
the vacuum energy belonging to the positive signature degrees of
freedom. In the usual quantum field theory vacuum energy is infinite
(or given by the Planck scale cutoff). Since any form of energy is
coupled to gravity, as a result we obtain the cosmological constant which is
drastically too high (120 orders of magnitude) in comparison with the
experimentally observed value. In our theory with vanishing vacuum energy,
there is no cosmological constant problem. In ref.\ci{PseudoHarm}
it is shown why such cancellation of vacuum energies is not in disagreement
with the Casimir effect and other effects of vacuum.
In our field theory fermions are automatically present as the elements of
the minimal ideals of ${\cal C}_{V_n}$. In the usual theory fermions are
included in order to satisfy the requirement of supersymmetry.
In supersymmetric field theory similar cancellation of vacuum energies occurs.

In the quantum field theory discussed here only the space of field
$V_n$ has been generalized to Clifford space. A next step is to generalize
spacetime, in which those fields live, as well, and consider a full
quantum field theory in $C$-space. A generalization to string field theory
also remains to be investigated.

In ref.\ci{Castro-Pavsic} curved $C$-space was considered and it was
found out that the Einstein-Hilbert action, generalized to $C$-space,
contained the ordinary higher derivative gravity in 4-dimensional
spacetime. Curved $C$-space also appears very promising for the
unification of fundamental interactions because it automatically
provides the required extra dimensions which in Kaluza-Klein theories
are postulated ad hoc and added to the known four dimensions of
spacetime. Following our approach, we can remain with four dimensions
of spacetime. Once we have a 4-dimensional spacetime, the 16-dimensional
Clifford space is automatically there: It is constructed from spacetime 
by means of its basis vectors $\gamma_\mu$ which are generators of
Clifford algebra. We only need to employ Clifford space for formulation
of physical theories. A part of this project is described in this paper,
more in refs. \ci{Pezzaglia}--\ci{CliffConform},\ci{Castro-Pavsic}, and
much more remains to be done.

\end{document}